# Traffic Instabilities in Self-organized Pedestrian Crowds

Mehdi Moussaïd[1,2,3*], Elsa G. Guillot[1,2], Mathieu Moreau[1,2], Jérôme Fehrenbach[4,5], Olivier Chabiron[4,5], Samuel Lemercier[6], Julien Pettré[6], Cécile Appert-Rolland[7,8], Pierre Degond[4,5] and Guy Theraulaz[1,2]

[1] Centre de Recherches sur la Cognition Animale, UMR-CNRS 5169, Université Paul Sabatier, Bât 4R3, 118 Route de Narbonne, 31062 Toulouse cedex 9, France.

[2] CNRS, Centre de Recherches sur la Cognition Animale, F-31062 Toulouse, France

[3] Center for Adaptive Behavior and Cognition, Max Planck Institute for Human Development, Lentzeallee 94, 14195 Berlin, Germany

[4] Institut de Mathématiques de Toulouse UMR 5219 (CNRS-UPS-INSA-UT1-UT2), Université Paul Sabatier, 118, route de Narbonne, 31062 Toulouse cedex, France

[5] CNRS, Institut de Mathématiques de Toulouse UMR 5219, 31062 Toulouse, France

[6] INRIA Rennes - Bretagne Atlantique, Campus de Beaulieu, 35042 Rennes, France

[7] Laboratoire de Physique Théorique, Université Paris Sud, bâtiment 210, 91405 Orsay cedex, France

[8] CNRS, UMR 8627, Laboratoire de physique théorique, 91405 Orsay, France

[*] Corresponding author: mehdi.moussaid@gmail.com

## *Classification*

Social Sciences/Psychological and Cognitive Sciences

Physical Sciences/Applied Physical Sciences

**Abstract**

In human crowds as well as in many animal societies, local interactions among individuals often give rise to self-organized collective organizations that offer functional benefits to the group. For instance, flows of pedestrians moving in opposite direction spontaneously segregate into lanes of uniform walking directions. This phenomenon is often referred to as a smart collective pattern, as it increases the traffic efficiency with no need of external control. However, the functional benefits of this emergent organization have never been experimentally measured, and the underlying behavioral mechanisms are poorly understood. In this work, we have studied this phenomenon under controlled laboratory conditions. We found that the traffic segregation exhibits structural instabilities characterized by the alternation of organized and disorganized states, where the lifetime of well-organized clusters of pedestrians follow a stretched exponential relaxation process. Further analysis show that the inter-pedestrian variability of comfortable walking speeds is a key variable at the origin of the observed traffic perturbations. We show that the collective benefit of the emerging pattern is maximized when all pedestrians walk at the average speed of the group. In practice, however, local interactions between slow- and fast-walking pedestrians trigger global breakdowns of organization, which reduce the collective and the individual payoff provided by the traffic segregation. This work is step ahead toward the understanding of traffic self-organization in crowds, which turns out to be modulated by complex behavioral mechanisms that do not always maximize the group's benefits. The quantitative understanding of crowd behaviors opens the way for designing bottom-up management strategies bound to promote the emergence of efficient collective behaviors in crowds.

**Author Summary**

A crowd of pedestrian is a complex system that exhibits a rich variety of self-organized collective behaviours. For instance, when two flows of people are walking in opposite directions in a crowded street, pedestrians spontaneously share the available space by forming lanes of uniform walking directions. This "pedestrian highway" is a typical example of self-organized functional pattern, as it increases the traffic efficiency with no need of external control. In this work, we have conducted a series of laboratory experiments to determine the behavioral mechanisms underlying this pattern. In contrast to previous theoretical predictions, we found that the traffic organization actually alternates in time between well-organized and disorganized states. Our results demonstrate that this unstable dynamics is due to interactions between people walking faster and slower than the average speed of the crowd. While the traffic efficiency is maximized when everybody walk at the same speed, crowd heterogeneity reduces the collective benefits provided by the traffic segregation. This work is a step ahead in understanding the mechanisms of crowd self-organization, and opens the way for the elaboration of management strategies bound to promote smart collective behaviors.

## *Introduction*

In many biological and social systems, such as fish schools, ant colonies, or human crowds, repeated local interactions among individuals support the emergence of a variety of collective patterns of motion [1,2,3,4]. Under certain conditions, the emerging organization allows the group to solve efficiently coordination problems without centralized planning or external control. In human crowds, such functional patterns of motion have been identified many times in the past, such as the alternating flows at a bottleneck [5], the formation of trails [6], or the walking configuration of social groups [7]. Remarkably, nobody orchestrates these phenomena and pedestrians do not actively seek these emerging collective organizations. Instead, individuals behave according to their own motivations, but local interactions generate functional organizations at the scale of the crowd. Therefore, these phenomena are often considered as prime examples of collective intelligence, sometimes called "the wisdom of crowds" [8,9,10].

One of the well-known example of such functional self-organization in crowds is the formation of lanes in bidirectional flows [11,12,13]: When two flows of pedestrians are moving in opposite directions in a crowded street, people spontaneously *share* the available space by forming a "pedestrian highway", where individuals walking in opposite direction segregate into lanes. This self-organized pattern of motion enhances the traffic flow by reducing frictional effects, local accelerations, energy consumptions and walking delays [14].

According to previous modeling works, the formation of lanes goes along with a sudden transition from disorder (where individuals are randomly distributed) to order (where opposite flows are segregated) [3]. Similar transitions from disorder to order have been observed in a wide variety of complex systems composed of locally interacting agents, in physical [15,16], biological [2,17,18,19] and social systems [10,20,21]. In human crowds, however, little is known about this phenomenon. From an empirical and quantitative point of view, the features of the spontaneous traffic organization remain scarcely documented, and the behavioral mechanisms underlying this phenomenon are hardly understood. In fact, it is poorly known how the transition operates, how the traffic organization evolves in time, and how much this collective organization benefits to the group. In this work, we have investigated the dynamics of lane formation under laboratory conditions, and studied the benefits provided by this traffic organization at the individual and crowd levels. To study the formation of lanes under experimental conditions, one major issue arising from past works is to handle the participants' inflow without interfering with the phenomenon [13,22]. In fact, in a straight corridor, the starting positions of pedestrians regularly introduced from both ends strongly influence the resulting traffic organization, which is detrimental to the relevance of the measurements. To avoid this drawback, we have used a ring-shaped corridor that provides periodic boundary conditions [18,23]. In this way, observing the phenomenon without perturbations induced by the experimental procedure becomes possible. To observe the emergence and the temporal dynamics of lane formation, N participants were randomly distributed in the ring-shaped corridor. A walking direction was randomly attributed to each of them, in such a way that N/2 participants walked clockwise and N/2 anti-clockwise. At the starting signal, participants started to walk in their attributed direction, allowing us to observe and characterize the emergence of traffic organization. A total of 11 replications were analyzed, with N=60, 50 and 30 participants (3, 2 and 6 replications, respectively).

In the following, we present our experimental results and show that the complex dynamics of traffic self-organization is based on simple behavioral mechanisms, where interactions between pedestrians walking faster and slower than the average trigger local perturbations that rapidly change into global traffic instabilities. While lane formation can be theoretically very efficient and functional, we show that, in practice, inter-individual variability undermines the overall benefits of the collective organization.

## *Results*

**Experimental results.** Our experimental results reveal a rapid transition from disorder to order during the first moments of the experiment, where initially randomly located pedestrians self-organize into lanes of opposite walking direction (**Fig.1**). However, the ordered state displays instabilities, where the flow segregation vanishes after a certain time lap and reappears again later and so on. In order to characterize this unstable dynamics, we have elaborated a clustering method to identify groups of pedestrians walking in lanes [24] (see Material and Methods). For this, we assume that a pedestrian *j* belongs to the same cluster as a pedestrian *i* if during a time period $\tau$, *j* passes at a distance smaller than $\delta$ from the position of *i* at time *t* (see the sketch

Fig.2). It appears that the number of clusters decreases rapidly after the beginning of the experiment, but displays alternating phases of order (i.e. five clusters or less) and disorder (i.e. ten clusters or more) (**Fig. 3**). To get a quantitative estimation of the traffic instability, we have measured the lifetime distribution of the clusters, where a cluster is considered as 'dead' when its composition changes by at least one individual. **Fig. 4A** shows the probability $p_i(t)$ for a cluster $i$ to be alive $t$ seconds after it appeared. As it can be seen, $p_i$ decays very fast during the first 10 seconds of a cluster lifetime. Yet, some clusters remain stable for 30 seconds or more. **Fig. 4B-C** show that $p_i(t)$ decays slower than an exponential and faster than a power law. **Fig. 4D** shows that $p_i(t)$ can be fitted empirically to a stretched exponential relaxation law [25,26]: $p_i(t) = e^{(at^k+b)}$, where $a$ and $b$ are the relaxation parameters, and $k$ is the relaxation exponent. The lifetime $\tau_0$ of a cluster can then be estimated by measuring the time after which a cluster has 95% chances to be changed by solving the equation $p_i(\tau_0) = 1 - 0.95$. Here, a numerical calculation gives $\tau_0$ =12.7($\pm$0.1), 8.4($\pm$0.2) and 7.8($\pm$0.2) seconds for N=30, 50 and 60 pedestrians, respectively.

Having characterized the typical time scale of the traffic instabilities, we will now investigate the origin of this dynamics: What are the behavioral mechanisms underlying these instabilities? Further analysis of our data reveals important density fluctuations in the experimental corridor, where highly crowded zones and almost empty zones can be observed at the same moment of time in different areas of the corridor (see **Video S3**). **Fig. 5A** shows empirically measured density maps, representing the local density value $\rho(\theta,t)$ for all times $t$ and in all directions $\theta$, as defined in the Material and Methods section. The density maps illustrate the spontaneous emergence of density gaps and density peaks that propagate along the corridor. Moreover, we have measured in a similar way the local radial speed $v_n(\theta,t)$ of traffic organization, which measures the lateral movements of pedestrians (see Material and Methods for a formal definition). In other words, $v_n(\theta,t)$ increases when pedestrians tend to move away from their lane, while it is close to zero when they walk one behind another. As shown in Fig. 5B, the place and time where the largest values of $v_n(\theta,t)$ occur coincide with the emergence of density gaps. In fact, Fig. 5C shows that the local radial speed is negatively correlated with the local density (a correlation test yields a p-value<0.01 with a correlation coefficient c=0.3, for all replications with N=60 pedestrians, and after removing the first 10 seconds of the experiments). What is the origin of these density fluctuations, and why are they correlated with important lateral movements? First, density gaps can be interpreted as a consequence of the variability in the comfortable walking speed: as pedestrians do not walk exactly at the same speed, those moving faster catch up with those walking slower, leaving an empty zone in front of the slow walkers. Second, the occurrence of lateral movements around the density gap can be explained in a similar way: pedestrians who are willing to walk faster than others make use of density gaps to overtake the slow walkers in front of them. By doing so, faster pedestrians move away from their lane, and meet the opposite flow head-on a few seconds later. This initial perturbation often triggers a complex sequence of avoidance maneuvers that results in the observed global instabilities. Therefore, we hypothesize that traffic instabilities result from the pedestrians walking speed variability, where people walking slowly unintentionally *create* density gaps, and those walking fast *make use* of these gaps to overtake their neighbors, triggering a chain reaction that results in the observed traffic instabilities.

**Computer simulations.** In order to validate this hypothesis and better understand the system dynamics, we have conducted a series of computer simulations under the same experimental conditions. To investigate the effects of the inter-individual variability of walking speeds, the comfortable speed $v_i^0$ of simulated pedestrians is randomly chosen at the beginning of each simulation according to a Gaussian distribution with mean $\bar{v}_0 = 1.2$ m/s and standard deviation $\sigma$ that varies from 0 (i.e. homogenous crowd) to 0.3 (i.e. large inter-individual differences). These values were chosen consistently with our experimental results, where the control tests indicate that the participants comfortable walking speeds are normally distributed with mean $\bar{v}_0 = 1.2$ m/s and standard deviation $\sigma_0 = 0.16$ (Kolmogorov-Smirnov test: p-value=0.73) (**Fig. 6**). Three examples of the dynamics observed during computer simulations are shown in **Fig. 7**.

By applying the clustering method defined above, we found that the clusters lifetime of simulated pedestrians also follow a stretched exponential relaxation law (**Fig. 4.D**). In particular, the relaxation exponents found in simulation are in good agreement with the experimentally determined ones. Furthermore, the simulation results indicate that the characteristic timescale of order phases decreases with increasing speed variability $\sigma$ (**Fig. 8.A)**. Therefore, this supports our hypothesis that speed variability is responsible for the observed traffic instabilities. Interestingly, this unstable dynamics is very likely to reduce the overall benefits of lane formation. Therefore, we used the model to measure the collective and the individual benefits of the flow segregation with increasing heterogeneity in the crowd. For this, we measured the collective payoff $\beta$ of the traffic organization by comparing the actual traffic flow of pedestrians to the average value measured when the N pedestrians move in the same direction: $\beta = (|Q_+| + |Q_-|)/Q_0$, where $Q_+$, and $Q_-$ are the average flow of pedestrians moving in clockwise direction, anti-clockwise direction in the bidirectional situation, and $Q_0$ is the average flow in unidirectional situation at the same density level. Therefore, $\beta = 1$ when pedestrians reach a collective organization that minimize the friction effects due to the opposite flows, providing the same traffic quality as a unidirectional situation. While a homogeneous crowd maximizes the collective payoff by forming stable lanes, the occurrence of traffic instabilities for higher values of $\sigma$ notably reduces the quality of the traffic flow (**Fig. 8.B**). We also measured the individual payoff of a pedestrian $i$: $P_i = (v_i \cdot e_i^0)/v_i^0$, which reflects how much the pedestrian approaches its desired speed $v_i^0$ and desired direction $e_i^0$. As shown **Fig. 8.C**, pedestrians who try to walk faster than the average have the lowest individual payoff, while those walking slower have the highest level of satisfaction. However, their combined effects have an important influence on the overall dynamics. In fact, even pedestrians who "cooperate" (i.e. those who have a desired speed close to the average) are increasingly less satisfied as the crowd becomes more heterogeneous, due to the strongest traffic instabilities induced by those who do not cooperate.

## *Discussion*

The spontaneous traffic organization of pedestrian flows is a functional self-organized collective pattern in human crowds, where people spontaneously share the available space by forming lanes of uniform walking direction. Based on experimental measurements, we found that this phenomenon exhibits structural instabilities, where mixed and well-segregated phases alternate in time. Our study demonstrates that speed variability among individuals is a key element underlying the observed traffic perturbations. Previous modeling work have suggested a similar relation between traffic stability and the fluctuations or the heterogeneity of the system, but these results were based on numerical simulations only [27,28]. In particular, our data allowed us to unravel the precise mechanisms underlying the emergence of traffic perturbations: people walking slower create density gaps, while those walking faster make use of these gaps to overtake other pedestrians in front of them. These specific local interactions finally result in large-scale traffic breakdowns, and the spontaneous self-organization ends up in a sub-optimal state. Therefore, the *collective* payoff of the group is undermined because pedestrians try to increase their *individual* level of satisfaction. Indeed, it is known that walking at the comfortable walking speed provides individual metabolic-related benefits [29]. But even pedestrians who cooperate by walking at the average group speed are increasingly less satisfied as other individuals deviating from the average speed are numerous. This incompatibility between individuals' satisfaction and crowd payoff is typical of many social dilemmas where self-interest conflicts with group interest [30,31].

Nevertheless, the functional benefit of traffic segregation is maximized in homogeneous crowds. Only diversity reduces the efficiency of the spatial self-organization. Many other decentralized systems facing coordination problems display the same trend. In car traffic, the variability among drivers' behaviors also lead to disturbing collective patterns, such as stop-and-go waves and large-scale traffic jams [9,32,33]. In other biological systems such as animal swarms, goal oriented collective motion is also disturbed by the presence of inter-individual variability [18,34]. Remarkably, when facing other kinds of tasks, inter-individual variability may have the opposite effect and promote the emergence of efficient behaviors [35,36,37]. In collective decision-making problems, heterogeneity favors the discovery of new solutions and prevents the group from staying stuck in suboptimal behaviors [38,39]. Therefore, it seems that group diversity can either promote or disturb collective intelligence depending on the nature of the task.

Among the rich variety of self-organized collective behaviors observed in human crowds, not all of them offer functional benefits to the group. While some phenomena like traffic segregation, or alternating flows at bottlenecks provide decentralized solutions to deal with congestion situations, other collective behaviors such as stop-and-go waves or crowd turbulence lead to serious traffic perturbation that may have life-threatening effects [40]. Therefore, understanding the mechanisms underlying these collective behaviors would open the way for the design of bottom-up management strategies bound to promote smart collective behaviors and minimize the risks during mass events. Our results already suggest real-life applications to enhance traffic efficiency and walking comfort in crowded walkways. For instance, dividing the pavement into a "fast lane" and a "slow lane" would reduce the overall speed variability in the crowd, and therefore avoid the emergence of traffic breakdowns. This appears to be particularly suited to crowded pedestrian walkways in large cities, where local commuters often meet up with foreign tourists. In the future, insights about pedestrian crowds may also serve as a basis for the investigation of other kinds of crowds, such as groups of web users, traders at stock market, or consumers [21,26,30,31].

## *Material and Methods*

**(a) Experimental design**. Controlled experiments were conducted in May 2009 at the INRIA in Rennes, France. The goal was to observe the emergence of spontaneous traffic organization in bidirectional flows of walking pedestrians. A total of 119 participants took part in the study, which conformed to the Declaration of Helsinki. They were naïve to the purpose of our experiments, and gave written and informed consent to the experimental procedure. None had known pathology that would affect their locomotion. Experiments were conducted in a ring-shaped corridor with inner radius $R_{in}$=2m and outer radius $R_{out}$=4.5m, providing a total surface of 51.05 $m^2$ (see **Fig.1**) built in a larger experimental room. As a control experiment, each participant was first instructed to walk alone in the experimental corridor (see **Fig. 6**). Then, we studied the effect of pedestrian density on the emergence of collective patterns of motion. Experimental trials were made with N=30, 50 and 60 pedestrians, corresponding to a global density level of 0.59, 0.98 and 1.18 p/$m^2$, respectively. A total of 3, 2 and 6 replications were reconstructed and analyzed for N=60, 50 and 30 participants, respectively. At the beginning of each trial, N participants were randomly distributed in the experimental corridor, and a walking direction was randomly attributed to each of them, in such a way that N/2 participants walked clockwise and N/2 anti-clockwise. At the starting signal, participants were asked to walk in their attributed direction as if they were moving alone in a street, and were not allowed to talk to each other (see **Video S1**). Each replication lasted for 60 seconds. The motion of each participant was recorded by means of an optoelectronic motion capture system (VICON MX-40, Oxford Metrics, UK). Participants were equipped with a white T-shirt and 4 reflexive markers, one on the forehead, one on the left acromion, and two on the right acromion to easily distinguish the left shoulder from right one. Markers motion was reconstructed using Vicon IQ software. The location of each participant was finally described as the center of mass of the 4 markers projected onto the horizontal plane (see **Video S2**).

**(b) Clustering method.** Two pedestrians belong to the same cluster at a given moment of time if one of them is following the other. We assume that a pedestrian *j* is following another pedestrian *i* at time *t*, if the trajectory of *j* in the time segment $[\ t\ \ t+\tau\ ]$ passes at a distance smaller than $\delta$ from the location of pedestrian *i* at time *t*. This definition of the clustering method is illustrated in **Fig.2.** The distance threshold was set to $\delta = 0.7m$ and the time window length was set to $\tau = 1s$. In the supporting information it is shown that the parameter values do not significantly affect the clustering outcome, as long as these lie in a reasonable interval (see **Text S1**, **Fig. S1** and **Fig. S2**).

**(c) Simulation model.** Simulations were performed by means of the previously published heuristics-based model for pedestrian behavior [3]. The model describes the adaptation of the actual velocity $\vec{v}_i$ of pedestrian *i* at time *t* by the acceleration equation $d\vec{v}_i/dt = (\vec{v}_{des} - \vec{v}_i)/\tau$, where $\tau$ is the relaxation time of 0.5 seconds, and the vector $\vec{v}_{des}$ is the desired velocity pointing in direction $\alpha_{des}$ and has the norm $\|v_{des}\| = v_{des}$. The desired direction $\alpha_{des}$ is given by minimizing the distance $d(\alpha)$ to the destination:

$$d^2(\alpha) = d_{max}^2 + f(\alpha)^2 - 2d_{max}f(\alpha)\cos(\alpha_0 - \alpha),$$

where $\alpha_0$ is the direction of the destination point $O_i$ and the function $f(\alpha)$ is the distance to the first collision if pedestrian *i* moved in direction $\alpha$ at his comfortable walking speed $v_i^0$, taking into account the other pedestrians' walking speeds and body sizes. For simplicity, we represent the pedestrian's body by a circle of radius $R_i$. If no collision is expected to occur in direction $\alpha$, $f(\alpha)$ is set to a default maximum value $d_{max}$, which represents the "horizon distance" of pedestrian *i*. The direction $\alpha$ is bounded by the vision field of the pedestrian, which ranges to the left and to the right by $\phi$ degrees with respect to the looking direction $\vec{H}_i$.

The desired velocity is given by the equation $v_{des}(t) = \min(v_i^0, d_h/\tau)$, where $d_h$ is the distance between pedestrian *i* and the first obstacle in the desired direction $\alpha_{des}$ at time *t*.

In cases of overcrowding, physical interactions between bodies may occur, causing unintentional movements that are not determined by the above heuristics. Therefore, in situations where the pedestrian *i* would be in physical contacts with other pedestrians, a repulsive force is used instead $\vec{f}_{ij} = kg(R_i + R_j - d_{ij})\vec{n}_{ij}$, where $g(x)$ is zero if the pedestrians $i$ and $j$ do not touch each other, and otherwise equals the argument $x$. $\vec{n}_{ij}$ is the normalized vector pointing from pedestrian $j$ to $i$, and $d_{ij}$ is the distance between the pedestrians' centers of mass. The physical interaction with a wall $W$ is represented analogously by a contact force $\vec{f}_{iW} = kg(R_i - d_{iW})\vec{n}_{iW}$, where $d_{iW}$ is the distance to the wall $W$ and $\vec{n}_{iW}$ is the direction perpendicular to it. Here again, the contact force with walls vanishes when the pedestrian does not touch the wall. The resulting acceleration equation then reads $d\vec{v}_i/dt = \sum_j \vec{f}_{ij}/m_i + \sum_W \vec{f}_{iW}/m_i$ and is solved together with the usual equation of motion $d\vec{x}_i/dt = \vec{v}_i$, where $\vec{x}_i(t)$ denotes the location of pedestrian $i$ at time $t$.

In order to simulate the movement of a pedestrian turning in the ring-shaped corridor, the destination point $Oi$ is updated at each simulation time step and located at a distance $d_O=5$ meters away in the direction tangent to the ring radius. The value of $d_O$ has been determined based on the control experiment results, by varying $d_O$ from 3m to 10m and choosing the value that minimizes the deviation between observed and predicted trajectories. The simulation parameters are $\tau=0.5$s, $\phi=45°$, $d_{max}=10$m, $k=10^3$, $R_i=0.2$m.

**(d) Measurement functions.** The local density $\rho(\theta,t)$ at time $t$ and in direction $\theta$ is defined as the average value of the local density $\rho(\vec{x},t)$, for all points $\vec{x}$ of the corridor located along the direction $\theta$ (with a reasonable spatial resolution). The local density is defined according to Ref. [40] as $\rho(\vec{x},t) = \sum_j f(d_{jx})$, where $d_{jx}$ is the distance between the center of mass of pedestrian $j$ and location $\vec{x}$, and $f(d)$ is a Gaussian-based weight function $f(d) = \frac{1}{\pi R^2}\exp(-d^2/R^2)$ with $R=0.7$ a weight parameter.

The local radial speed $v_n(\theta,t)$ is defined as the average radial speed $\vec{n}_j$ of all pedestrians $j$ located between directions $\theta_1 = \theta - \Delta\theta$ and $\theta_2 = \theta + \Delta\theta$ at time $t$, where the parameter $\Delta\theta$ is set to $\pi/16$. The radial speed $\vec{n}_j$ is given by $\vec{n}_j = \Delta r_j/\Delta t$, where $r_j$ is the radial position of pedestrian $j$ in the experimental step.

## Acknowledgements

We are grateful to the M2S research group at the Université de Rennes 2 for their help and expertise during the experiments, and in particular to Armel Crétual, Richard Kulpa, Antoine Marin, and Anne-Hélène Olivier. We thank J.Gautrais, D.Helbing, J.Gouello and the two anonymous referees for inspiring comments on the work.

**References**

1. Helbing D, Molnar P, Farkas IJ, Bolay K (2001) Self-organizing pedestrian movement. Environ Plan B 28: 361-383.
2. Sumpter D (2010) Collective Animal Behavior. Princeton: Princeton University Press.
3. Moussaïd M, Helbing D, Theraulaz G (2011) How simple rules determine pedestrian behavior and crowd disasters. Proc Natl Acad Sci USA 108: 6884-6888.
4. Ondrej J, Pettré J, Olivier A-H, Donikian S (2010) A synthetic-vision based steering approach for crowd simulation. ACM Transactions on Graphics 29: 1-9.
5. Kretz T, Wölki M, Schreckenberg M (2006) Characterizing correlations of flow oscillations at bottlenecks. J Stat Mech P02005.
6. Helbing D, Keltsch J, Molnar P (1997) Modelling the evolution of human trail systems. Nature 388: 47-50.
7. Moussaïd M, Perozo N, Garnier S, Helbing D, Theraulaz G (2010) The Walking Behaviour of Pedestrian Social Groups and Its Impact on Crowd Dynamics. PLoS ONE 5: e10047.
8. Helbing D, Johansson A (2009) Pedestrian, crowd, and evacuation dynamics. In: Meyers RA, editor. Encyclopedia of Complexity and Systems Science: Springer. pp. 6476-6495.
9. Surowiecki J (2004) The wisdom of crowds. New-York: Doubleday.
10. Ball P (2004) Critical Mass: How One Thing Leads to Another. New York: Farrar, Straus and Giroux.
11. Older SJ (1968) Movement of pedestrians on footways in shopping streets. Traffic Eng Contr 10: 160-163.
12. Milgram S, Toch H (1969) Collective Behavior: Crowds and Social Movements. In: Lindzey G, Aronson E, editors. Handbook of Social Psychology. pp. 507-610.
13. Kretz T, Grünebohm A, Kaufman M, Mazur F, Schreckenberg M (2006) Experimental study of pedestrian counterflow in a corridor. J Stat Mech P10001.
14. Helbing D, Vicsek T (1999) Optimal self-organization. New J Phys 1: 1-17.
15. Nicolis G, Prigogine I (1977) Self-Organization in Nonequilibrium Systems: From Dissipative Structures to Order through Fluctuations. New-York: John Wiley & Sons.
16. Deseigne J, Dauchot O, Chaté H (2010) Collective motion of vibrated polar disks. Phys Rev Lett 105: 098001.
17. Camazine S, Deneubourg J-L, Franks N, Sneyd J, Theraulaz G, et al. (2001) Self-Organization in Biological Systems. Princeton: Princeton University Press.
18. Buhl J, Sumpter DJT, Couzin ID, Hale JJ, Despland E, et al. (2006) From Disorder to Order in Marching Locusts. Science 312: 1402-1406.
19. Couzin I, Franks N (2003) Self-organized lane formation and optimized traffic flow in army ants. Proc R Soc B 270: 139-146.
20. Nèda Z, Ravasz E, Brechet Y, Vicsek T, Barabasi AL (2000) The sound of many hands clapping. Nature 403: 849-850.
21. Schelling T (1978) Micromotives and Macrobehavior. New York: WW Norton & Company.
22. Daamen W, Hoogendoorn S (2002) Controlled experiments to derive walking behaviour. Eur J Transp Infrast 3: 39-59.

23. Sugiyama Y, Fukui M, Kikuchi M, Hasebe K, Nakayama A, et al. (2008) Traffic jams without bottlenecks - Experimental evidence for the physical mechanism of the formation of a jam. New J Phys 10: 033001.
24. Vicsek T, Zafiris A (2010) Collective motion. Phys Rep. In press.
25. Laherrèe J, Sornette D (1998) Stretched exponential distributions in Nature and Economy: 'Fat tails' with characteristic scales. Eur Phys J B 2: 525-539.
26. Wu F, Huberman B (2007) Novelty and collective attention. Proc Natl Acad Sci USA 104: 17599-17601.
27. Burstedde C, Kirchner A, Klauck K, Schadschneider A, Zittartz J. (2002) Cellular Automaton Approach to Pedestrian Dynamics - Applications. In: Schreckenberg M, Sharma SD, editors. Pedestrian and Evacuation Dynamics: Springer. pp. 87-98.
28. Campanella M, Hoogendoorn S, Daamen W (2009) Effects of Heterogeneity on Self-Organized Pedestrian Flows. Trans Res Rec 2124: 148-156.
29. Donelan JM, Kram R, Kuo AD (2001) Mechanical and metabolic determinants of the preferred step width in human walking. Proc R Soc B 268: 1985-1992.
30. Glance N, Huberman B (1994) The Dynamics of Social Dilemmas. Sci Am 270: 76-81.
31. Olson M (1971) The Logic of Collective Action : Public Goods and the Theory of Groups: Harvard University Press.
32. Helbing D, Huberman B (1998) Coherent moving states in highway traffic. Nature 396: 738-740.
33. Kerner B (2004) The Physics of Traffic: Empirical Freeway Pattern Features, Engineering Applications, and Theory: Springer.
34. Vicsek T, Czirók A, Ben-Jacob E, Cohen I, Shochet O (1995) Novel Type of Phase Transition in a System of Self-Driven Particles. Phys Rev Lett 75: 1226-1229.
35. Bonabeau E, Theraulaz G (1999) Role and variability of response thresholds in the regulation of division of labor in insect societies. In: Detrain C, Deneubourg JL, Pasteels J, editors. Information Processing in Social Insects: Birkhauser Verlag. pp. 141-163.
36. Mattila H, Seeley T (2007) Genetic Diversity in Honey Bee Colonies Enhances Productivity and Fitness. Science 317: 362-364.
37. Pruitt J, Riechert S (2011) How within-group behavioural variation and task efficiency enhance fitness in a social group. Proc R Soc B 278: 1209-1215.
38. Couzin I, Krause J, Franks N, Levin S (2005) Effective leadership and decision-making in animal groups on the move. Nature 433: 513-516.
39. Couzin I (2008) Collective cognition in animal groups. Trends Cogn Sci 13: 36-43.
40. Helbing D, Johansson A, Al-Abideen H (2007) The Dynamics of crowd disasters: an empirical study. Phys Rev E 75: 046109.

**Figure Legends**

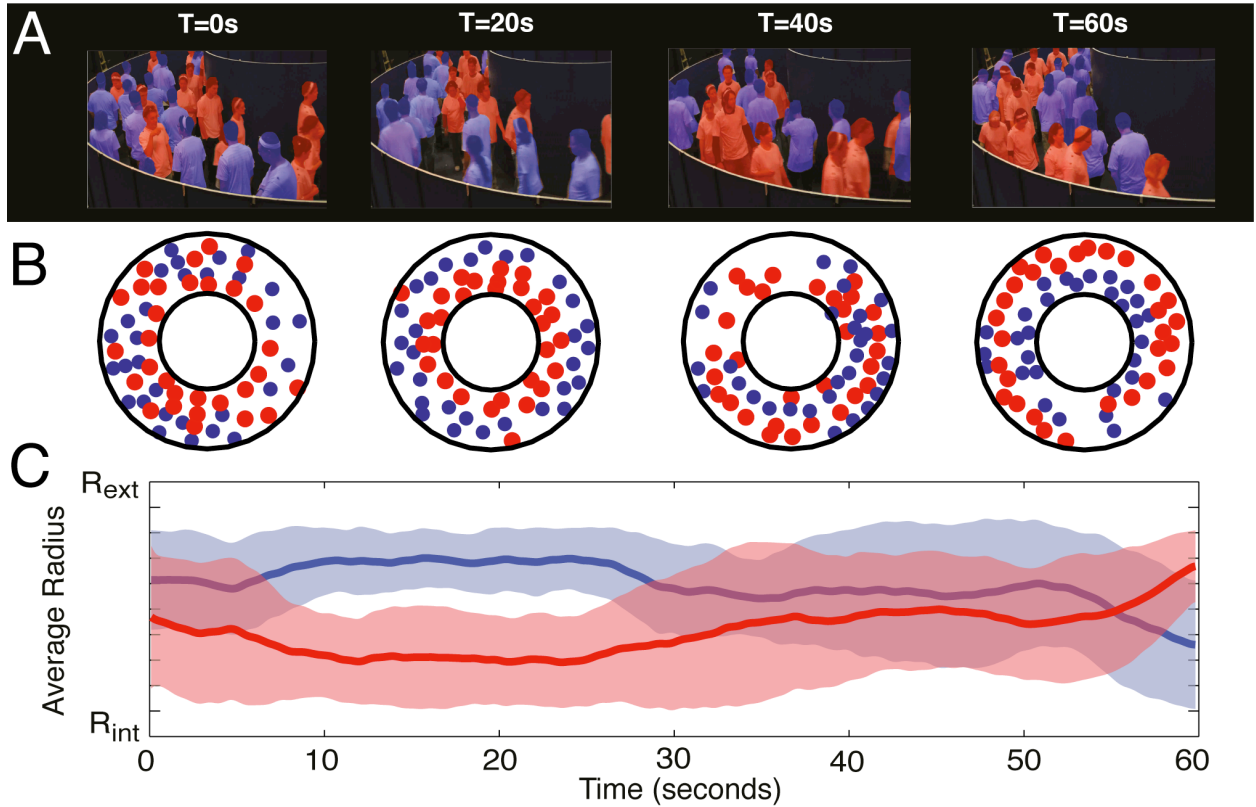

**Figure 1**: Illustration of the unstable dynamics observed under experimental conditions for one replication with N=60 pedestrians. Starting from a random initial state, the flows of pedestrians segregate after a short transition time (approximately 10 seconds in this replication). However, this ordered state turns to disorder after a certain period of time, before order emerges again later at the end of the experiment. (A) The upper figures are snapshots from the laboratory experiment, where people have been colored in red or blue according to their direction of motion. Blue pedestrians turn clockwise and red anti-clockwise. (B) The output of the tracking system with the same color-coding. (C) The average radial position $r_i$ of all pedestrians *i* moving in the same direction, where the same color-coding is used. $R_{int}$ and $R_{ext}$ denote the internal and the external walls of the ring-shaped corridor. The alternation of mixed and segregated phases is visible. The transparent areas show the standard deviation of the mean value.

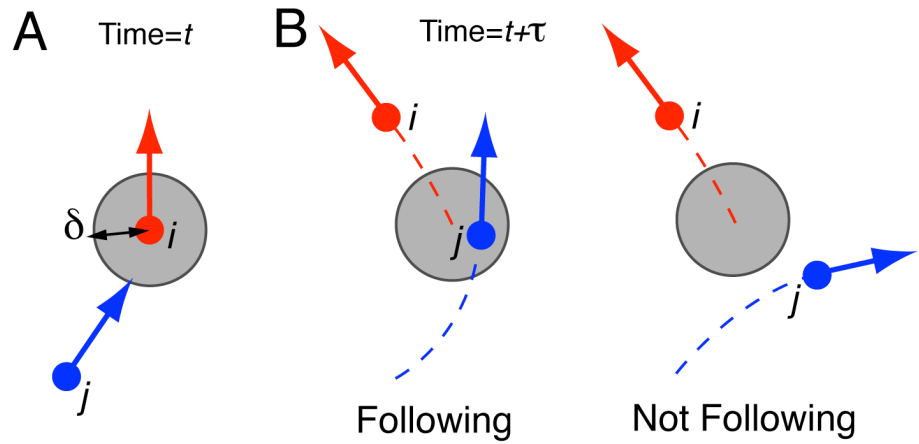

**Figure 2:** Illustration of the clustering method. (A) Two pedestrians $i$ and $j$ belong to the same cluster if one follows the other. (B) The pedestrian $j$ follows pedestrian $i$, if $j$ moves closer than a distance $\delta$ from the position of pedestrian $i$ at time $t$, during a time period of $\tau$ seconds. Here, $\tau=1$s and $\delta=0.6$m are two clustering parameters.

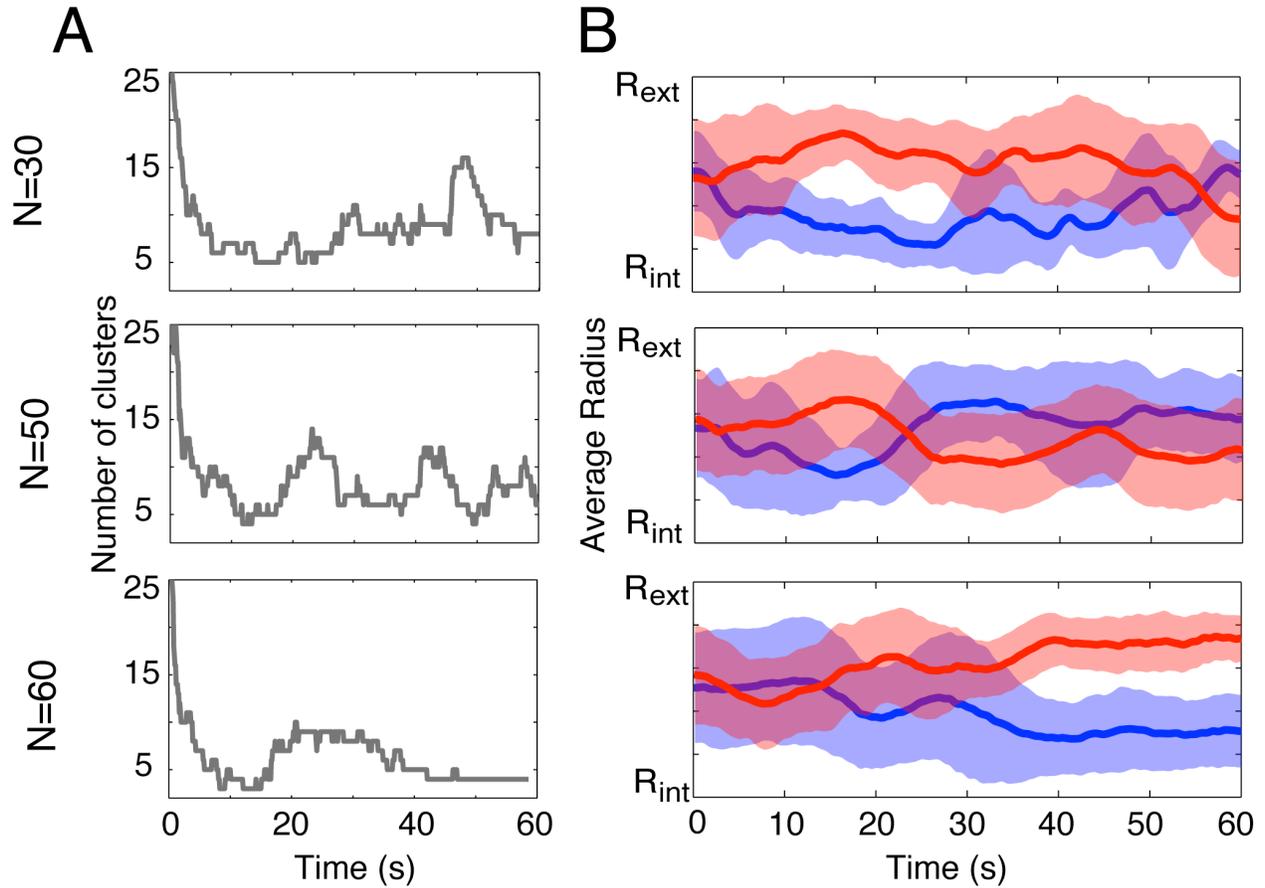

**Figure 3**: (A) Illustration of the evolution of the number of clusters for three replications with N=30, 50 and 60 pedestrians. The clustering method is described in the Material and Methods section and illustrated Fig. 2. During the first ten seconds, the initial transition from disorder to order is visible. Then, the number of clusters oscillates between well-organized (five clusters or less), and disorganized states (ten clusters or more). (B) The corresponding segregation dynamics for the same three replications.

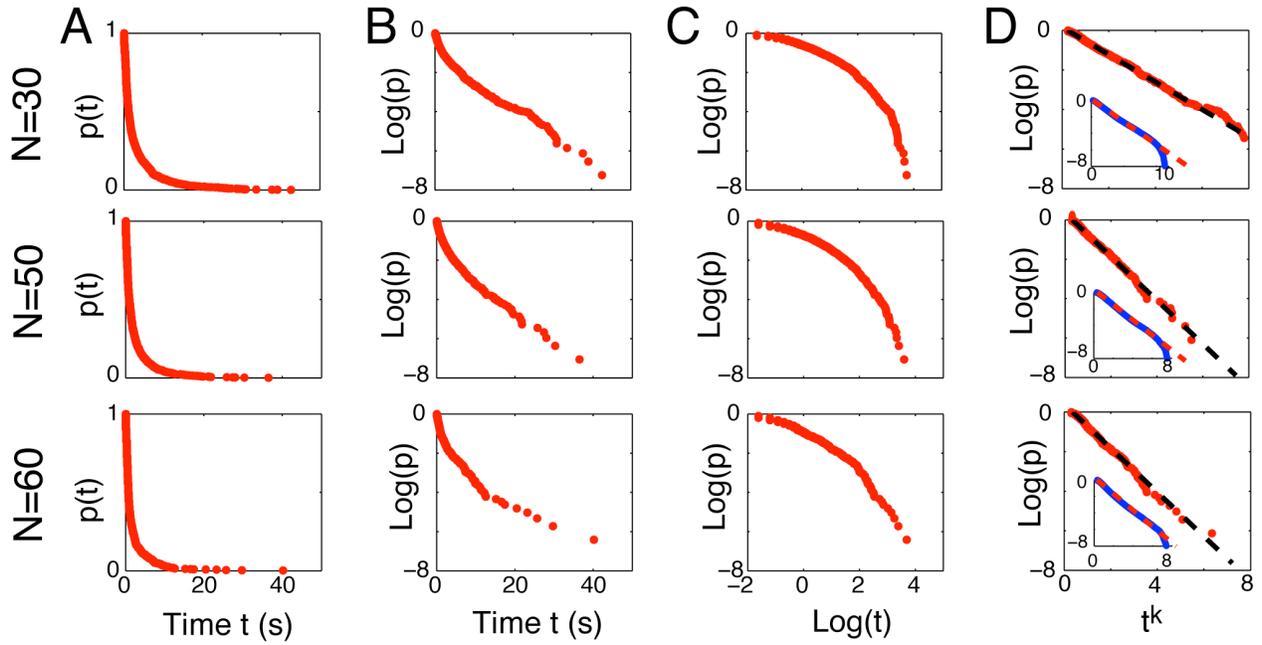

**Figure 4**: Empirical distribution of the clusters lifetime. (A) The probability $p(t)$ for a cluster to remain unchanged after a time period of $t$ seconds. (B) $log(p)$ versus $t$ does not yield a straight line, showing that $p(t)$ decays slower than an exponential. (C) $log(p)$ versus $log(t)$ is a curve, showing that $p(t)$ decays faster than a power-law. (D) A straight line is found for $log(p)$ versus $t^k$ with $k=0.4$, demonstrating that the lifetime of pedestrian clusters follows a stretched exponential relaxation law: $p_i(t) = e^{(at^k + b)}$, where the relaxation exponent $k$ depends on the number of pedestrians N. The insets indicate simulation results, where the same distribution law is found. Empirical data and computer simulations yield the same relaxation exponents $k$=0.6, 0.5, and 0.5 for N=30, 50 and 60 respectively.

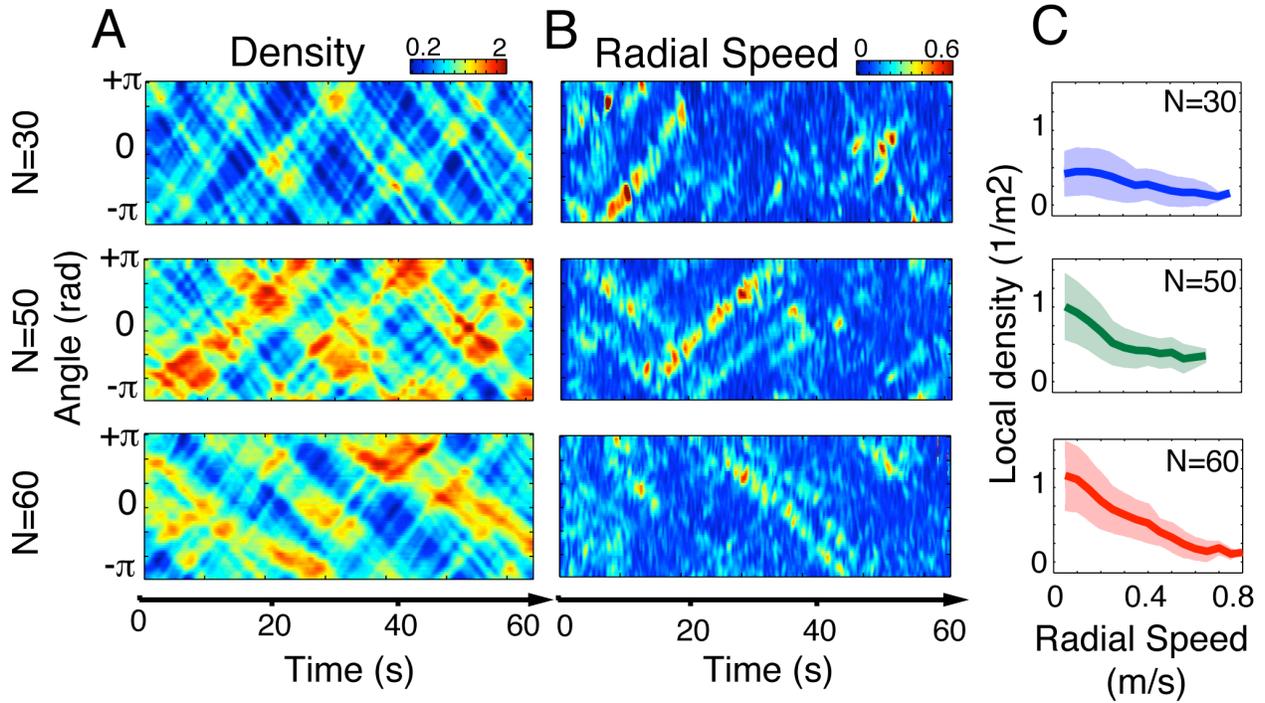

**Figure 5:** Correlation between local radial speed and density gaps. (A) Local density maps $\rho(\theta,t)$ for three representative replications with N=30, 50 and 60 pedestrians. The emergence and the propagation of density peaks (red) and density gaps (blue) are visible. (B) Local radial speed $v_n(\theta,t)$ for the same three replications, showing the lateral movements of pedestrians. The largest values occur mostly around density gaps. (C) Average local density as a function of local radial speed, for all replications with N=30, 50 and 60 pedestrians. The largest values of $v_n(\theta,t)$ occur where the local density level is low, that is, around density gaps. This correlation is less visible for N=30, probably due to the lower global density level.

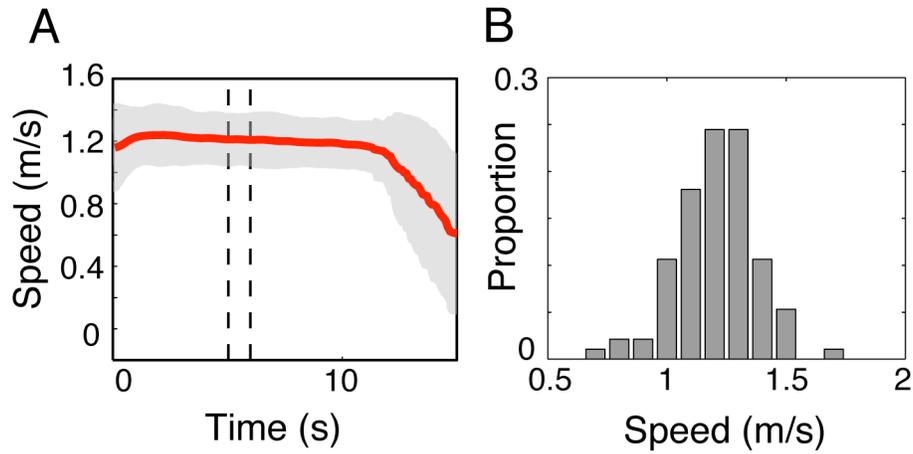

**Figure 6**: Characterization of the walking behaviour during the control test. (A) The average walking speed of all participants as they were walking alone in the experimental corridor. The grey area indicates the standard deviation of the mean. The dashed lines are the limits of the measurement zone, where the pedestrians are assumed to have reached their comfortable walking speed. (B) The comfortable walking speeds are normally distributed with mean $\bar{v}_0 = 1.2$ m/s and standard deviation $\sigma_0 = 0.16$ (a Kolmogorov-Smirnov test yields a p-value of 0.73).

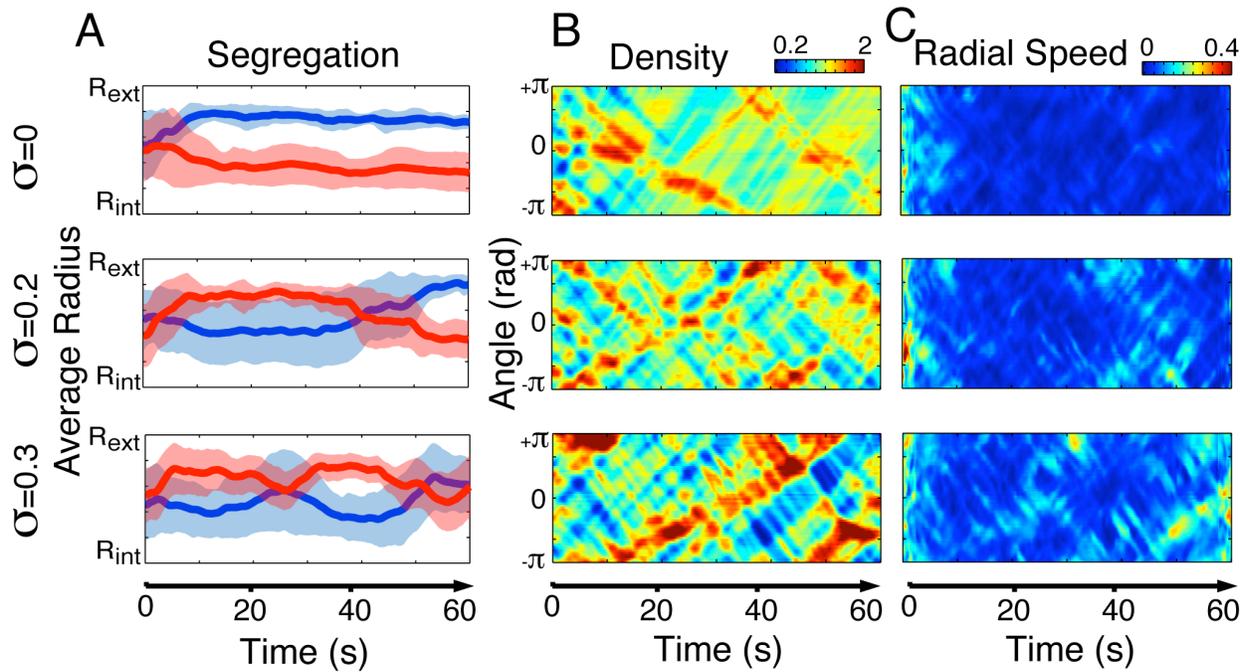

**Figure 7:** Illustration of the dynamics observed during computer simulations. (A) As speed variability increases from $\sigma=0$ to $\sigma=0.3$, the model predicts an increasingly unstable segregation dynamics. These instabilities go along with the emergence of increasingly sharp density gaps (B), which leads to stronger and more frequent lateral movements (C). The time and place where lateral movements occur in (C) fit with the propagation of density waves in (B) and explain the unstable dynamics observed in (A).

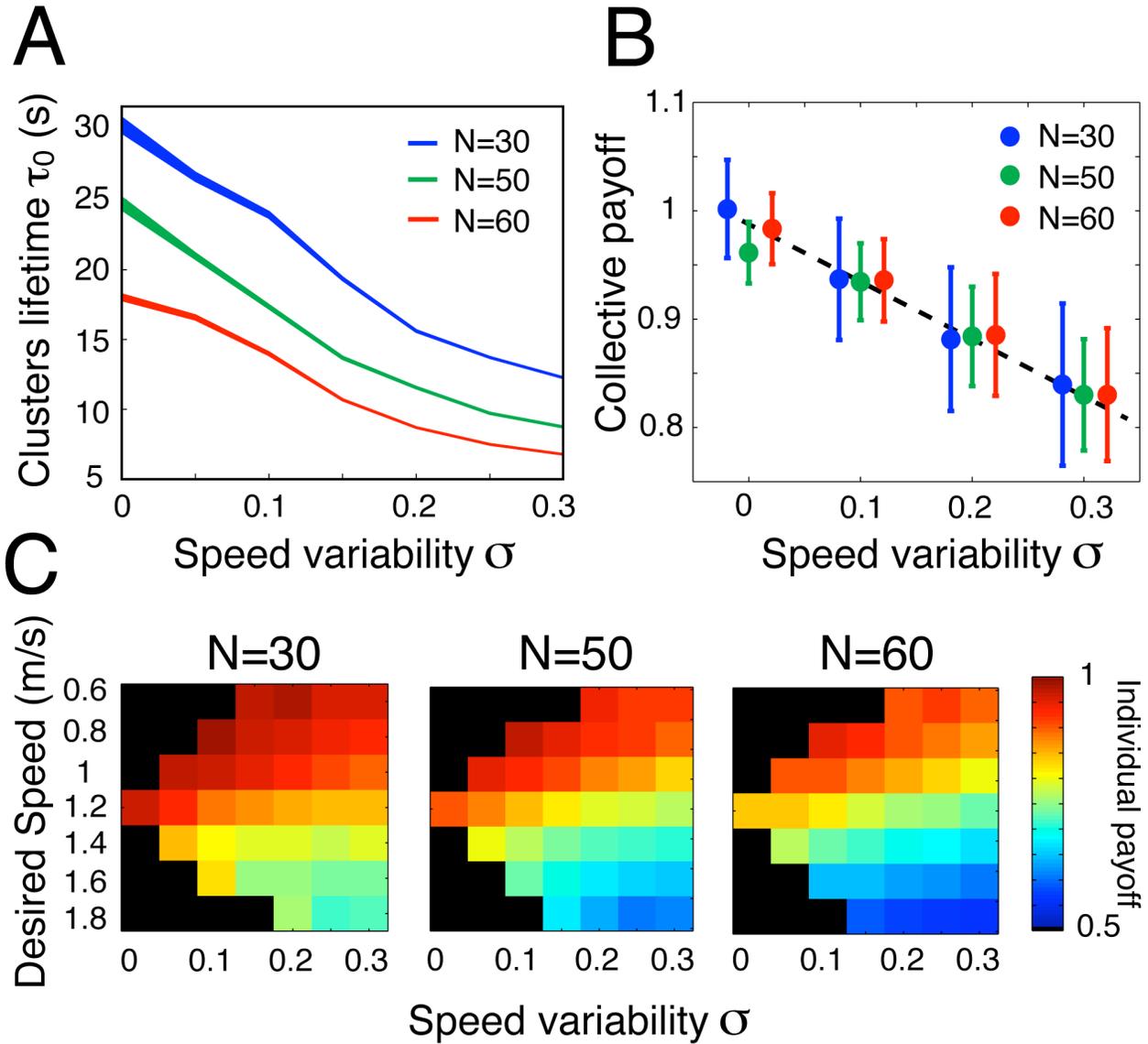

**Figure 8**: Collective dynamics predicted in simulations. (A) Cluster lifetime $\tau_0$ as a function of the standard deviation $\sigma$ of the comfortable walking speed distribution, as predicted by numerical simulations. The decreasing curves demonstrate the relationship between inter-individual variability and traffic instabilities. The width of the curves indicates the 95% confidence bounds of the lifetime estimation. (B) The collective payoff provided by the lane organization, as a function of $\sigma$. (C) The individual payoff of pedestrians averaged over all simulations for N=30, N=50, and N=60, grouped according to their desired walking speed. The black areas indicate the absence of value.

**Supplementary Information**

**Text S1:** Parametric sensitivity study for the clustering method.

**Figure S1:** Illustration of the outcome of the clustering technique for a replication with N=60 pedestrians, where a cluster number has been automatically attributed to each individual. A well-organized situation is shown on the left (4 clusters), and a disorganized state is shown on the right (11 clusters). Blue pedestrians turn clockwise and red pedestrians anti-clockwise.

**Figure S2:** Surface plot of the mean number of clusters detected for 50 pedestrians experiments (A), and 60 pedestrians experiments (B).

**Video S1:** Video recording of an experiment with N=60 pedestrians.

**Video S2:** The dataset for an experiment with N=60 participants, as obtained after the tracking and data reconstruction process.

**Video S3:** Illustration of the emerging density fluctuations, for an experiment with N=60 participants. The color-coding indicates the local density value ($1/m^2$).